\documentclass{PoS}
\usepackage{graphicx}

\def\x{\mathbf{x}}
\def\y{\mathbf{y}}
\def\colora{}
\def\colorb{}
\def\colorc{}

\title{The Early Stages of
Heavy Ion Collisions}

\ShortTitle{The Early Stages of
Heavy Ion Collisions}

\author{{Fran\c cois Gelis}\\
        Universit\'e Paris-Saclay, CNRS, CEA, Institut de physique th\'eorique\\ 91191, Gif-sur-Yvette, France\\
        E-mail: \email{francois.gelis@ipht.fr}}

\abstract{Heavy ion collisions pose interesting challenges to quantum
chromodynamics, because they probe the parton structure of the
incoming nuclei at very small longitudinal momentum fractions.
Combined with the large size of nuclei, this may lead to the
phenomenon of gluon saturation. The Color Glass Condensate is an
effective QCD description that aims to cope with such a situation. In
this talk, I show how one may study heavy ion collisions in this
framework.}

\FullConference{Light Cone 2019 - QCD on the light cone: from hadrons to heavy ions - LC2019\\
		16-20 September 2019\\
		Ecole Polytechnique, Palaiseau, France}

\begin{document}


Lattice QCD has taught us that nuclear matter should undergo a
transition at high temperature and/or density, by which the quarks and
gluons confined into hadrons in ordinary conditions become
deconfined. These conditions of temperature and density were most
certainly reached in the early Universe, but the deconfined nuclear
matter present at those times did not leave any visible imprint
accessible to present day astronomy. These critical conditions can
also be briefly accessed by colliding heavy nuclei at very high
energy, as performed by the RHIC and the LHC. 
\begin{figure}[htbp]
  \centering
  \includegraphics[width=4cm]{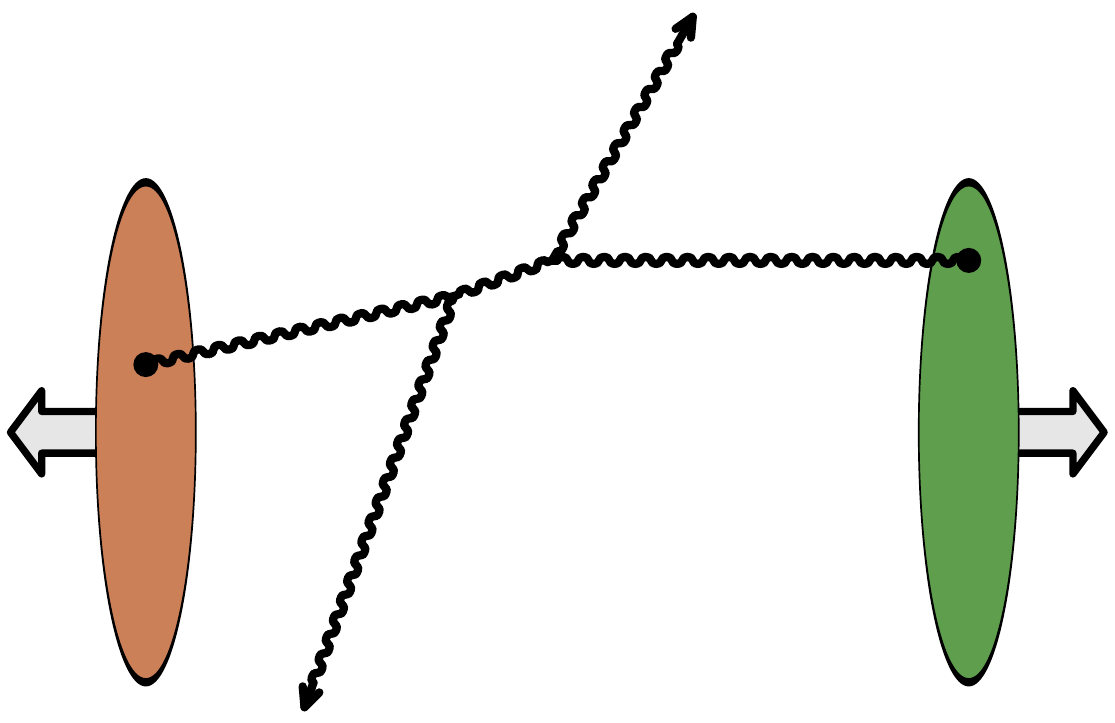}
  \hskip 5mm
  \includegraphics[width=4cm]{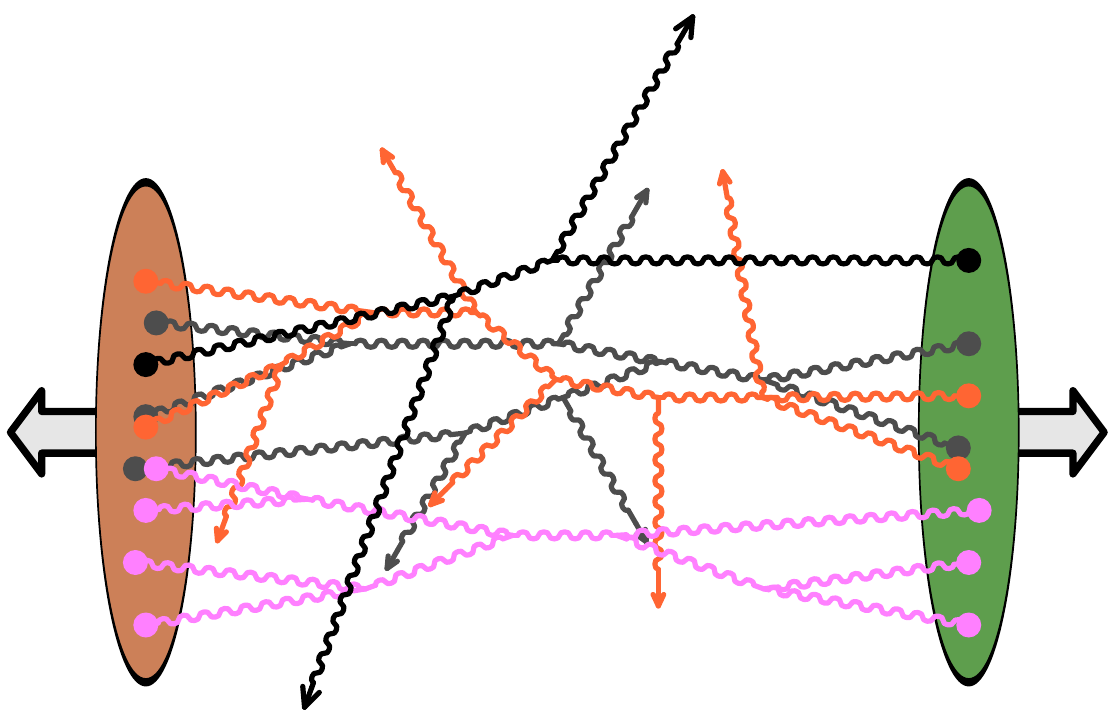}
  \hskip 10mm
  \includegraphics[width=4cm]{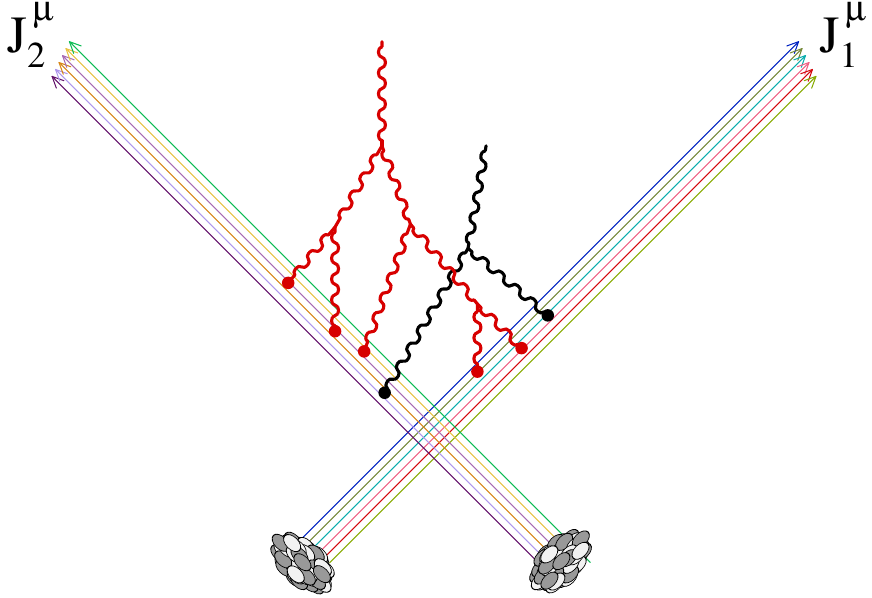}
 \caption{\label{fig:dense}Left: collisions of dilute and dense projectiles. Right: color currents and radiated gauge field.}
\end{figure}
From the point of view of QCD, the main difficulty in studying these
collisions is the fact that the nuclei are probed in a regime of large
gluon density. In fact, not only the standard formalism of parton
distributions becomes inadequate because it does not provide any
information regarding multi-gluon distributions, but also gluons in
this regime undergo non-linear interactions that limit the growth of
their density, a phenomenon known as gluon saturation \cite{Gribov:1984tu}. The color glass
condensate \cite{Gelis:2010nm} is an effective description of hadrons or nuclei in the
dense regime, in which the projectiles are treated as color currents
flowing along the light-cone. 
These currents are eikonally coupled to the gauge fields and become
strong (of order $g^{-1}$) in the saturated regime, which leads to a
peculiar power counting \cite{Gelis:2006yv}, \setbox1\hbox to
4cm{\includegraphics[width=4cm]{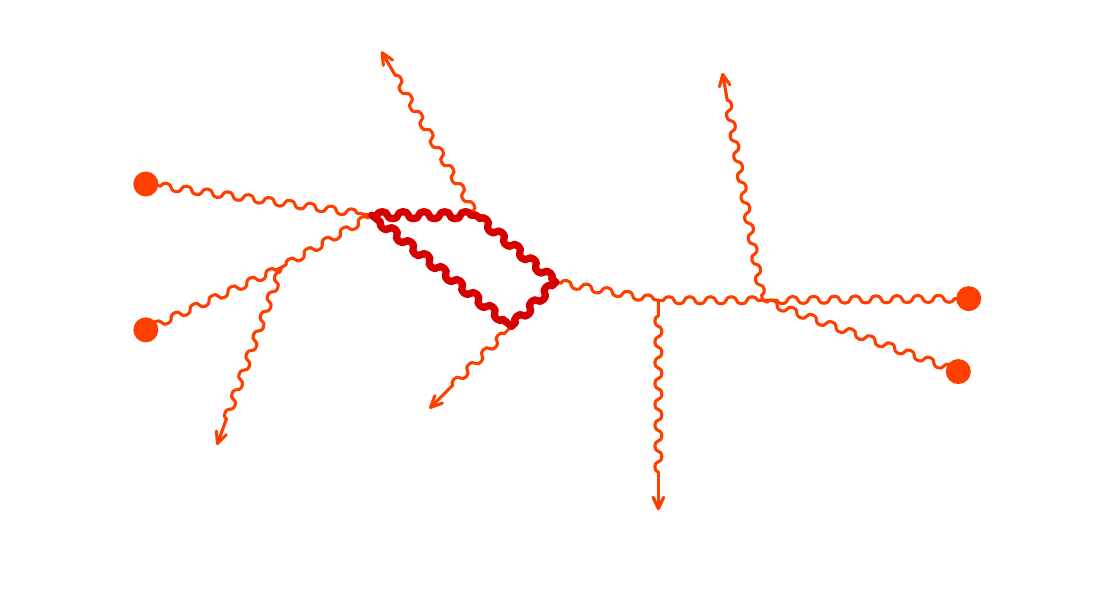}}
\begin{equation}
  \raise -13mm\box1\;\sim\quad
  \underbrace{g^{n_{_E}-2}}_{\mbox{\scriptsize ext. lines}}\; 
\underbrace{ (\hbar g^2)^{n_{_L}}}_{\mbox{\scriptsize loops}}\; 
\underbrace{(g\,J)^{n_j}}_{\mbox{\scriptsize sources}},
\end{equation}
where the number of insertions of the currents is irrelevant in the
saturated regime (each connected vacuum graph is of order $g^{-2}$). In general, observables in the saturated regime are
obtained as the sum of an infinite number of graphs, including both
graphs connected to the measured gluons and disconnected vacuum graphs. 
\begin{figure}[htbp]
  \centering
  \includegraphics[width=6cm]{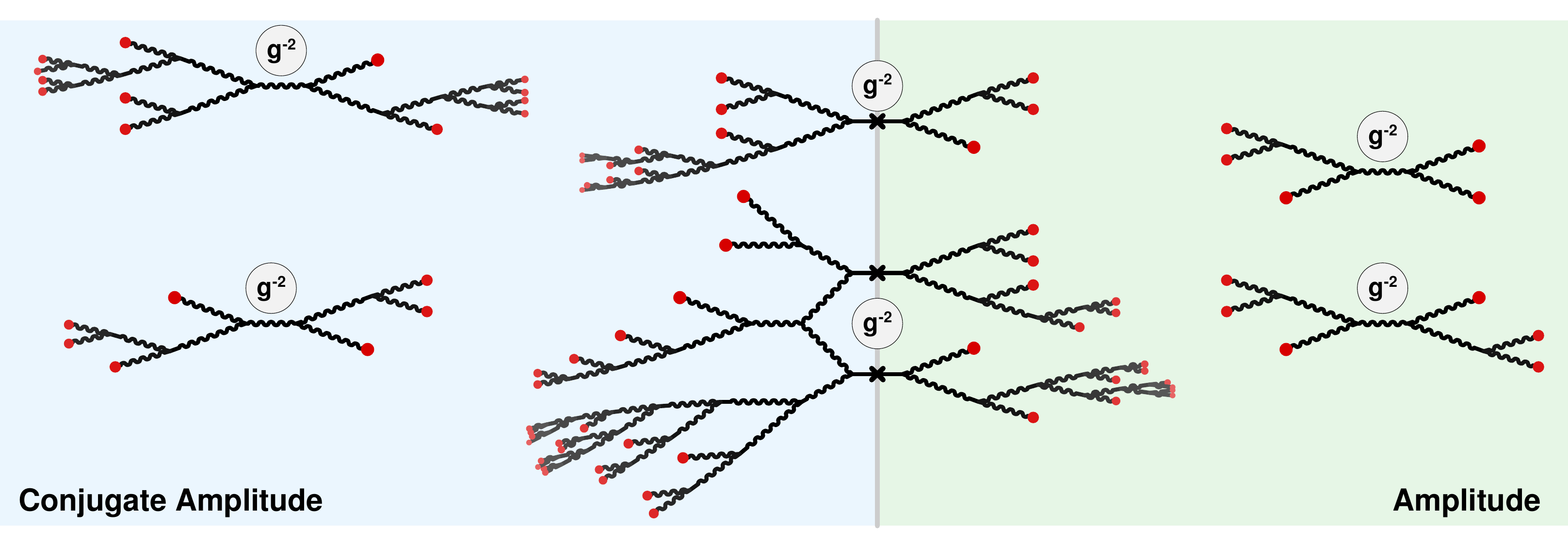}
  \caption{\label{fig:exclusive}Diagrammatic expansion for a generic observable.}
\end{figure}
Inclusive observables are considerably simpler since the disconnected
vacuum graphs drop out \cite{Gelis:2006yv}. For instance, gluon production is determined
at leading order uniquely by the sum of all the tree diagrams, i.e., by
the classical solution of Yang-Mills equations (moreover, one can show
that this solution must obey a null retarded boundary condition):
\setbox1\hbox to 1.6cm{\resizebox*{1.6cm}{!}{\includegraphics{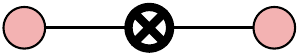}}}
\setbox2\hbox to
10cm{\resizebox*{10cm}{!}{\includegraphics{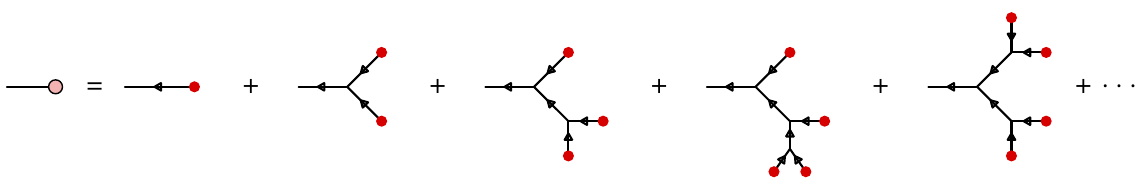}}}
\begin{equation}
  \overline{N}_{_{\rm LO}}=\quad\raise -0.7mm\box1,\quad
  \raise -7mm\box2
\end{equation}
An important question in the context of heavy ion collisions is that of the
thermalization of the produced gluonic matter. Immediately after the
collision, this matter is far off-shell, made of color fields whose
electric and magnetic components are parallel to the collision
axis \cite{Lappi:2006fp}. Consequently, the initial longitudinal pressure is exactly the
opposite of the energy density. By solving numerically the classical
Yang-Mills equations, one observes that the longitudinal pressure
increases and reaches positive values at a time $Q_s \tau \sim 1$, but
never becomes comparable to the transverse pressure (in fact, at
leading order, this system appears to be a free-streaming collection
of gluons) \cite{Krasnitz:1999wc,Fukushima:2011nq}.
\begin{figure}[htbp]
  \centering
  \includegraphics[width=7cm]{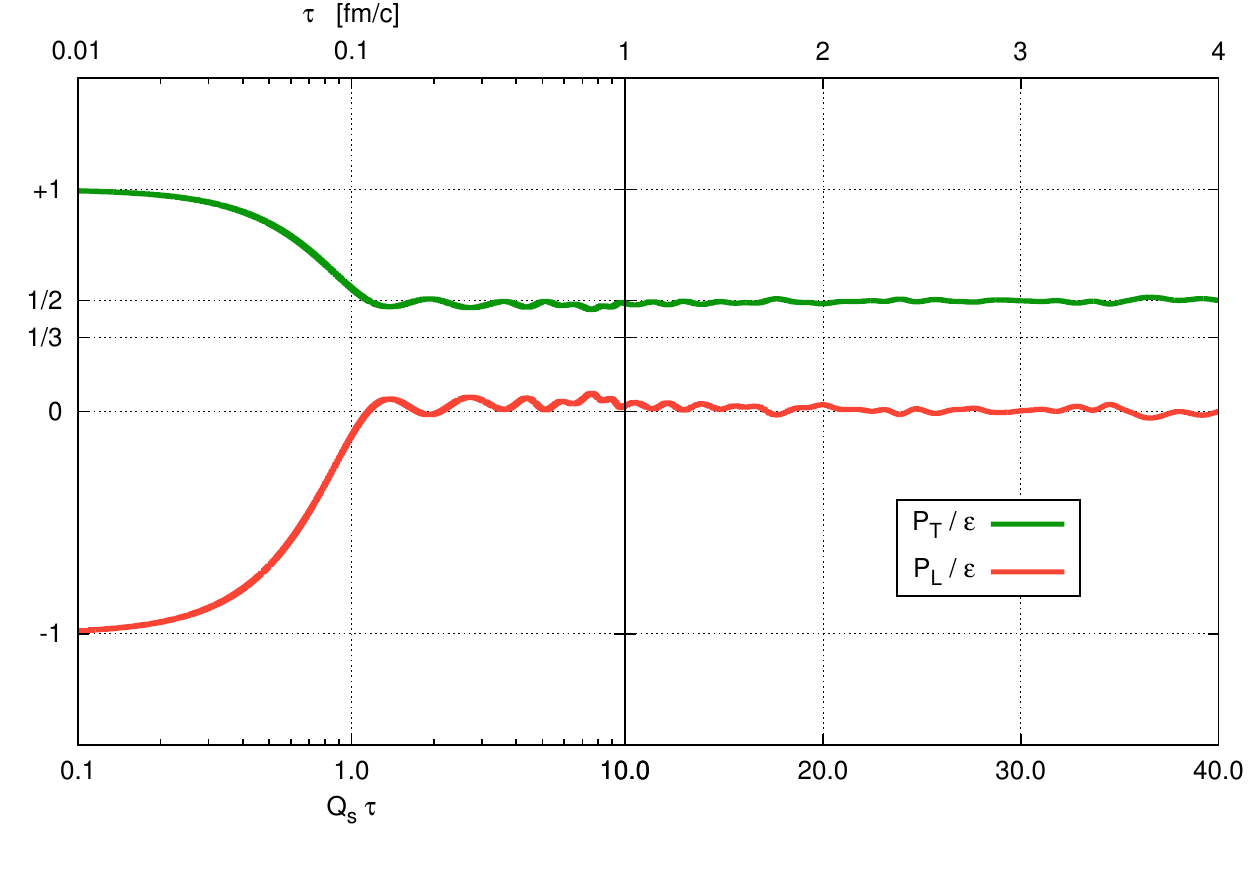}
  \caption{\label{fig:LO-Press}Time evolution of the transverse and
    longitudinal pressures at leading order.}
\end{figure}

The next order is made of all the one-loop graphs, in the presence of
the external currents representing the projectiles, as illustrated
here: \setbox1\hbox to
48mm{\resizebox*{48mm}{!}{\includegraphics{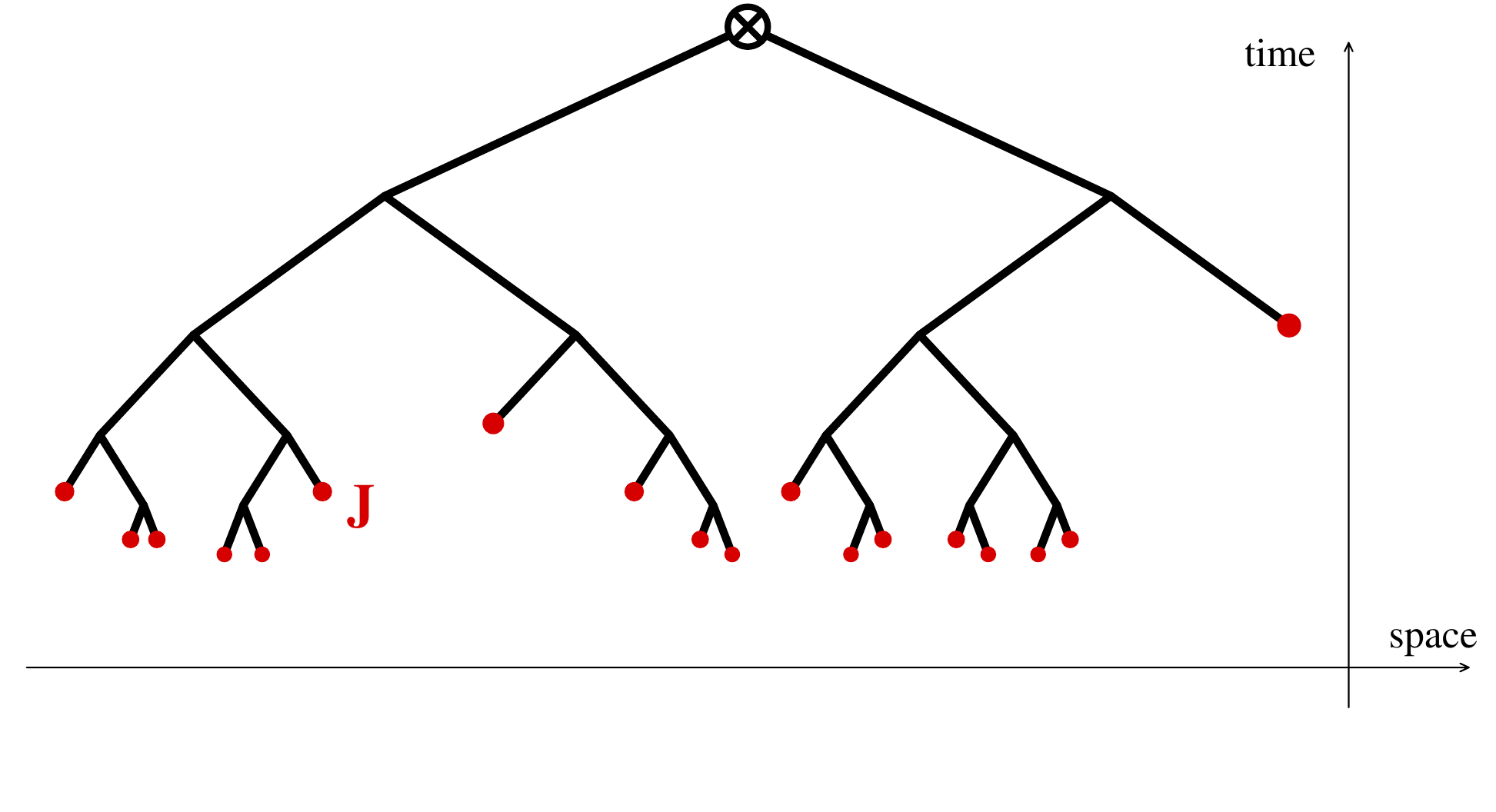}}} \setbox2\hbox
to 48mm{\resizebox*{48mm}{!}{\includegraphics{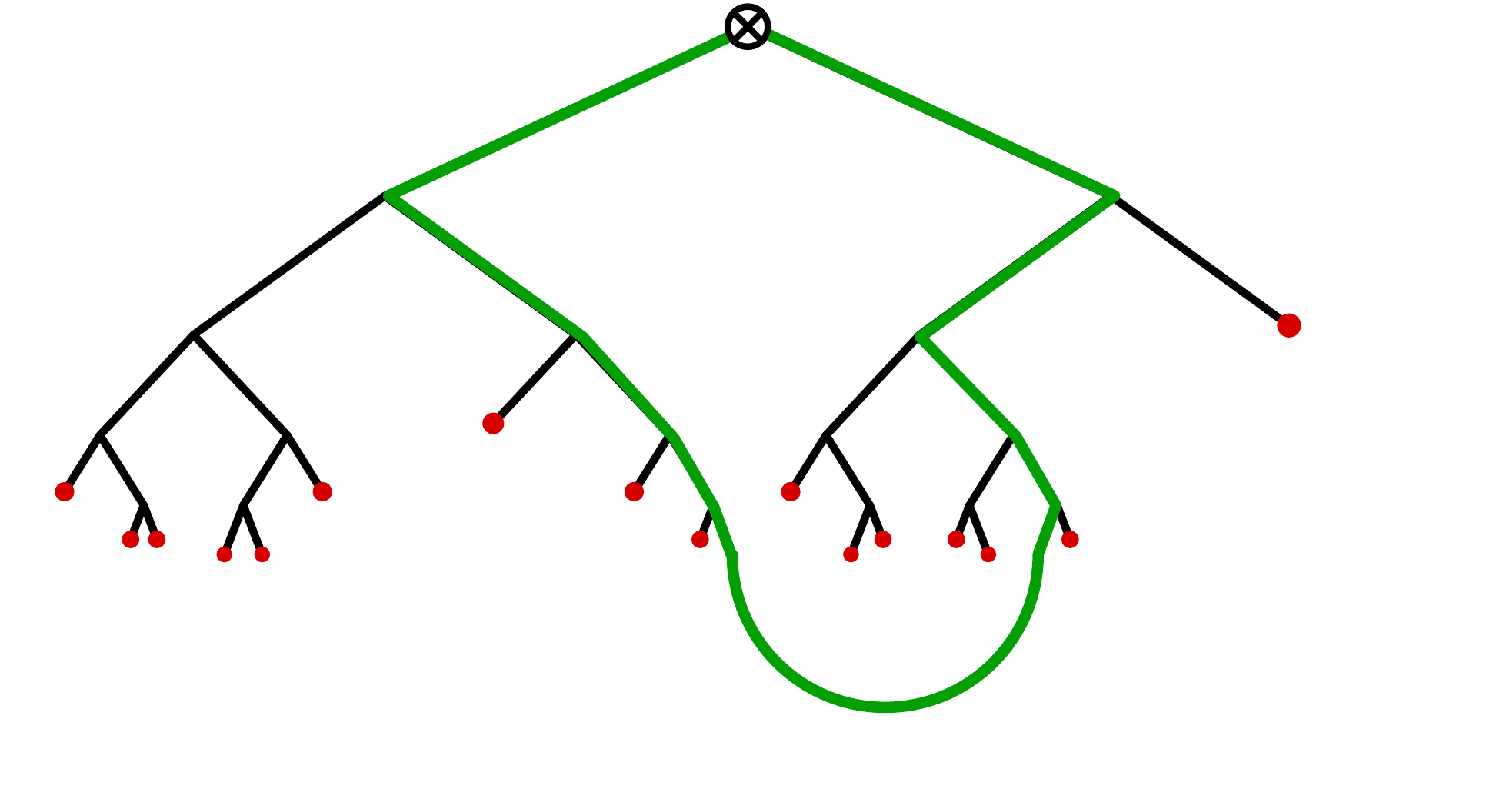}}}
\begin{equation}
{\cal O}_{_{\rm LO}}[J]=\raise -15mm\box1,\quad
{\cal O}_{_{\rm NLO}}[J]=\raise -15mm\box2
\end{equation}
In the case of inclusive observables, such a one-loop correction can
be related to the leading order contribution \cite{Gelis:2008rw}. The causal nature of the
classical field encountered at leading order plays a crucial role in
these manipulations. Loosely speaking, this amounts to first extending
the original observable so that it depends on a field with a non-null
initial condition, and then to note that the one-loop correction to
this generalized quantity can be obtained by the action of an operator
which is quadratic in derivatives with respect to the initial value of
the field. Diagrammatically, this relationship reads
\setbox1\hbox to 45mm{\resizebox*{45mm}{!}{\includegraphics{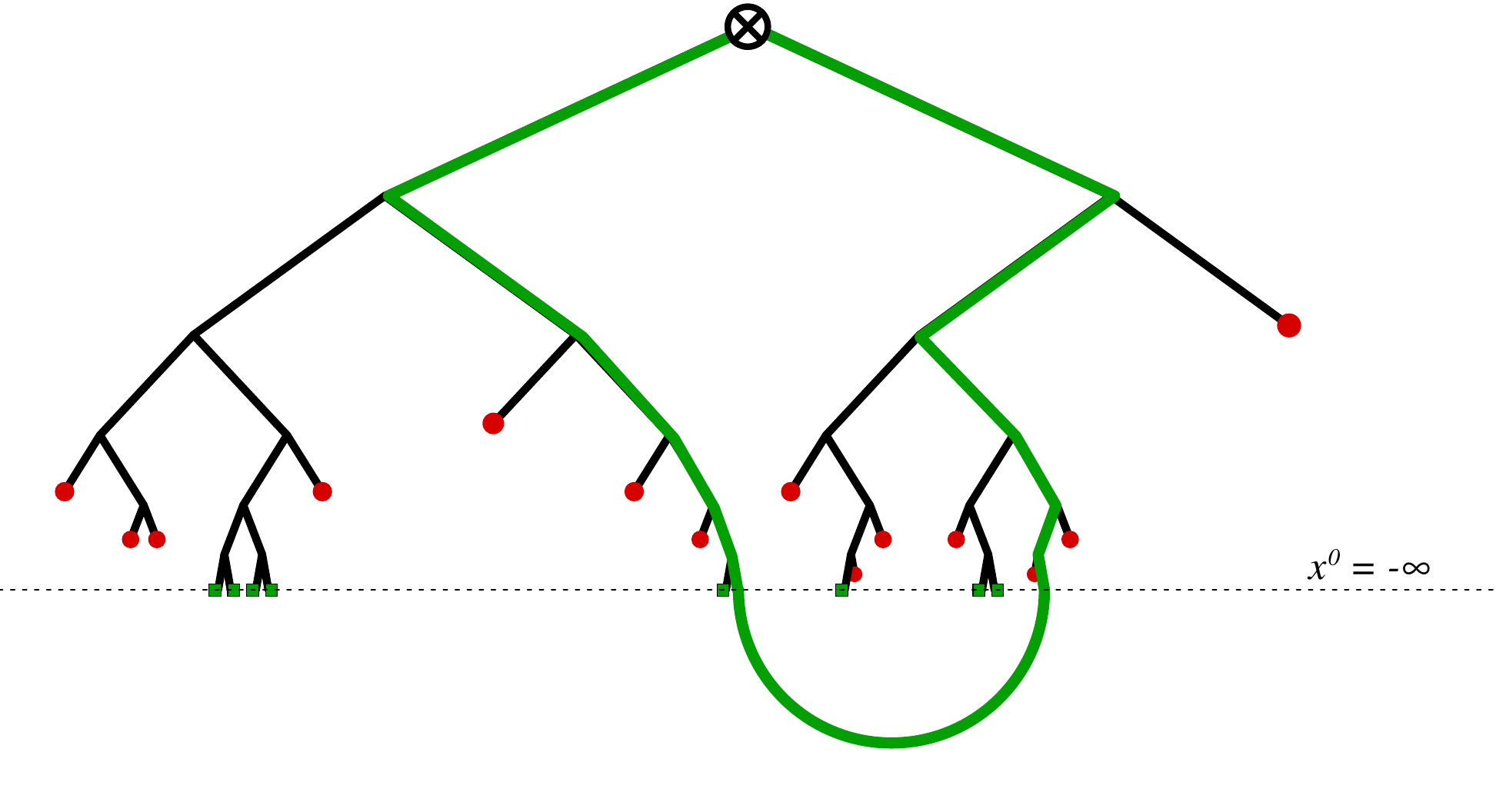}}}
\setbox2\hbox to 45mm{\resizebox*{45mm}{!}{\includegraphics{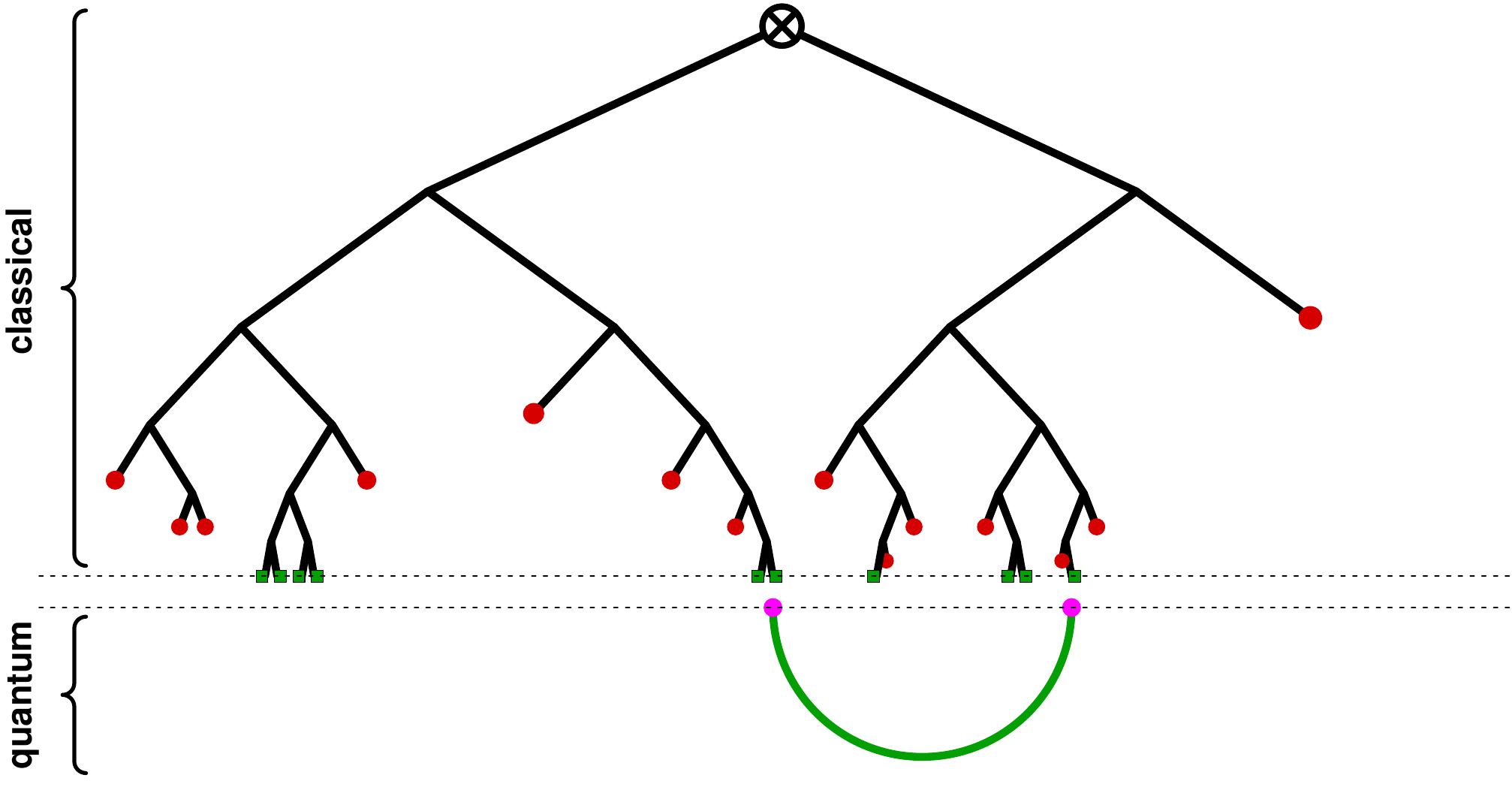}}}
\begin{equation}
{\cal O}_{_{\rm NLO}}[J,{\cal A}_{\rm in}]=\raise -15mm\box1,\quad
{\cal O}_{_{\rm NLO}}[J,{\cal A}_{\rm in}]=\raise -15mm\box2,
\end{equation}
while a more precise version is
\begin{equation}
{\cal O}_{_{\rm {NLO}}}[J,{\cal A}_{\rm in}]
=
       \Bigg[\frac{{\hbar}}{2}\int d^3\x d^3\y\;{{\Gamma}(\x,\y)}\;
              \frac{\delta}{\delta{\cal A}_{\rm in}(\x)}
              \frac{\delta}{\delta{\cal A}_{\rm in}(\y)}
              \Bigg]\;\vphantom{\frac{\delta}{\delta{\cal A}_{\rm in}(\x)}}{\cal O}_{_{\rm LO}}[J,{\cal A}_{\rm in}].
              \label{eq:nlo}
\end{equation}
In this equation, $\Gamma(\x,\y)$ is a two-point function at equal
times (both times being at $x^0 \to -\infty$).  This formula tells
that in order to obtain the NLO, one should take the LO, remove two
instances of the initial field and connect the handles thus freed by
the link $\Gamma(\x,\y)$ in order to form a loop.

\begin{figure}[htbp]
  \centering
  \includegraphics[width=5cm]{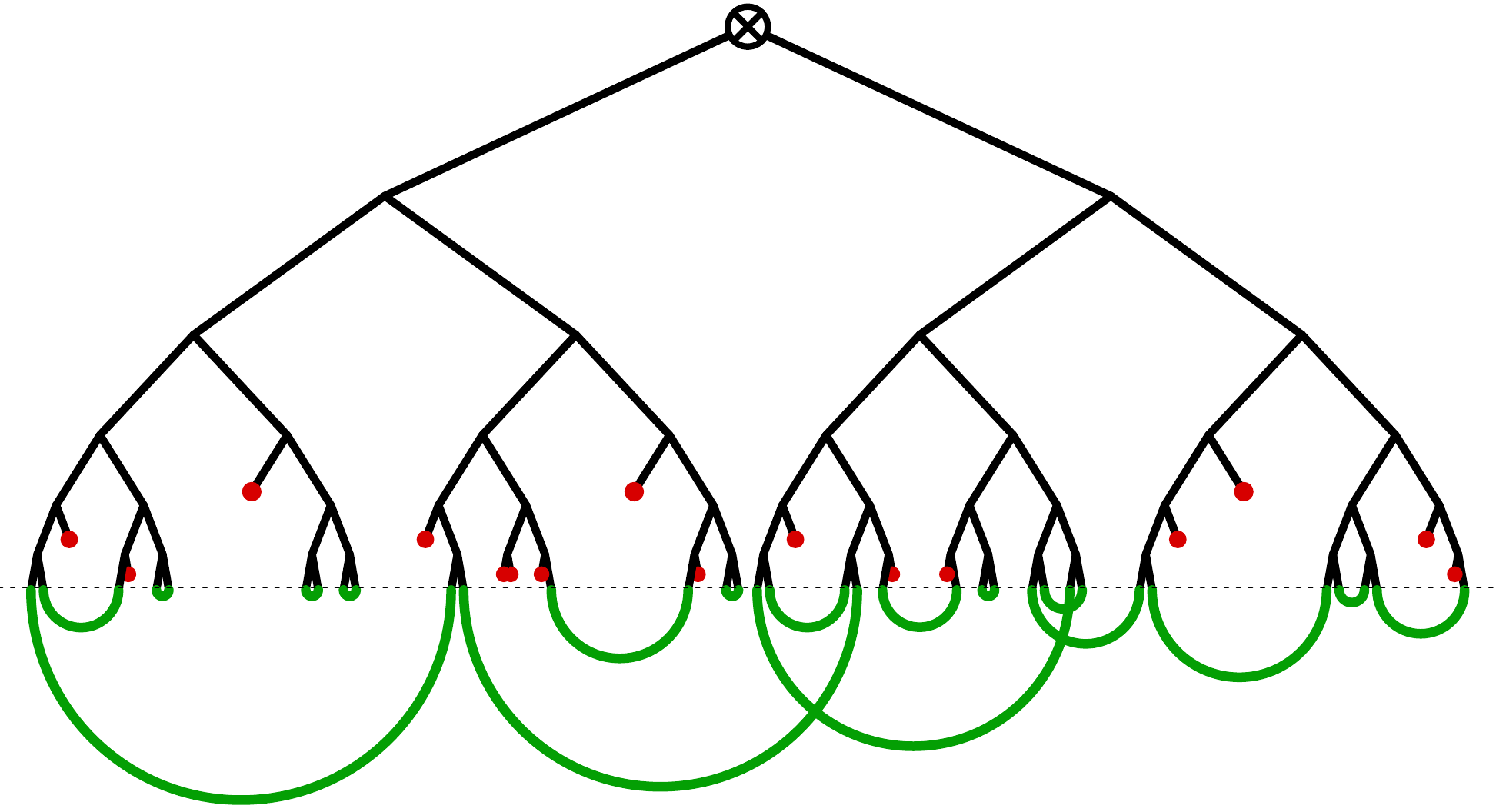}
  \caption{\label{fig:resum}Graphs with the fastest growth.}
\end{figure}

In the regime of strong sources, this manipulation brings an extra
factor $g^2$, as expected for a loop. However, this power counting is
upset in situations where the classical solutions are unstable. In
this case, perturbing the initial condition generally leads to an
exponential growth controlled by the Lyapunov exponent.
\begin{figure}[htbp]
  \centering
  \includegraphics[width=4.5cm]{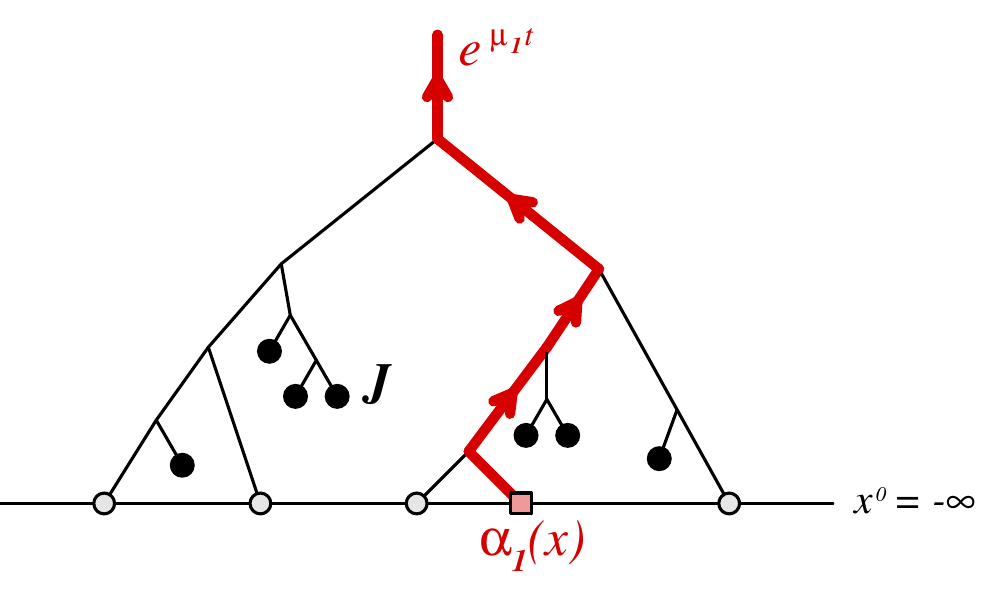}
  \hskip 5mm
  \includegraphics[width=4.5cm]{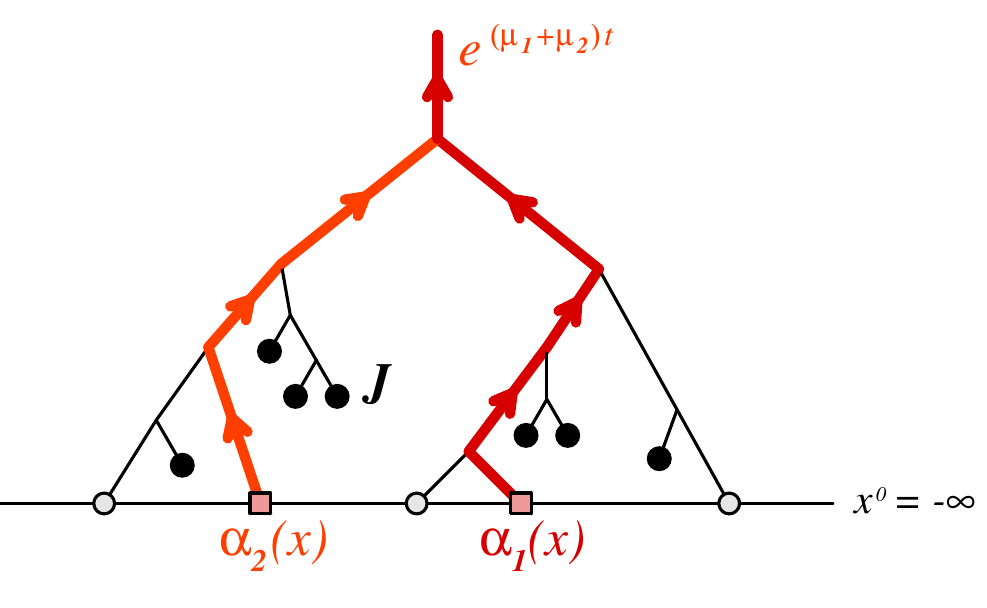}
  \caption{\label{fig:lyapunov}Exponential growth of perturbations to
    a classical solution.}
\end{figure}
Furthermore, this exponent is proportional to the number of points
where the initial condition is perturbed. Based on this, one can
easily see that the graphs with the fastest growth are those depicted
in Figure \ref{fig:resum}. 
The sum of all these graphs can be generated simply by exponentiating the
operator in eq. (\ref{eq:nlo}),
\begin{eqnarray}
{\cal O}_{_{\rm {resummed}}}[J,{\cal A}_{\rm in}]
&=&
       \exp\Bigg[\frac{{\hbar}}{2}\int d^3\x d^3\y\;{{\Gamma}(\x,\y)}\;
              \frac{\delta}{\delta{\cal A}_{\rm in}(\x)}
              \frac{\delta}{\delta{\cal A}_{\rm in}(\y)}
              \Bigg]\;\vphantom{\frac{\delta}{\delta{\cal A}_{\rm in}(\x)}}{\cal O}_{_{\rm LO}}[J,{\cal A}_{\rm in}]\nonumber\\
              &=&
\int \big[D{\colorc a}\big]\;\exp\Big[-\frac{1}{2\,{\colorb\hbar}}\int_{\x,\y}
{\colorc a(\x)}{\colora{\Gamma}^{-1}(\x,\y)}{\colorc a(\y)}\Big]\;{\cal O}_{_{\rm LO}}[{\cal A}_{\rm in}+{\colorc a}].
\end{eqnarray}
The second line is an exact alternate representation of the action of
this exponentiated operator, more suitable for a practical
implementation. Indeed, it indicates that this resummation can be
obtained as an average over classical solutions, obtained by
superimposing Gaussian fluctuations to the LO initial condition (this
resummation is known as the {\sl Classical Statistical
  Approximation}). The function $\Gamma$ is known both in scalar
theory and in Yang-Mills theory \cite{Epelbaum:2013waa}.
\begin{figure}[htbp]
  \centering
  \includegraphics[width=5cm]{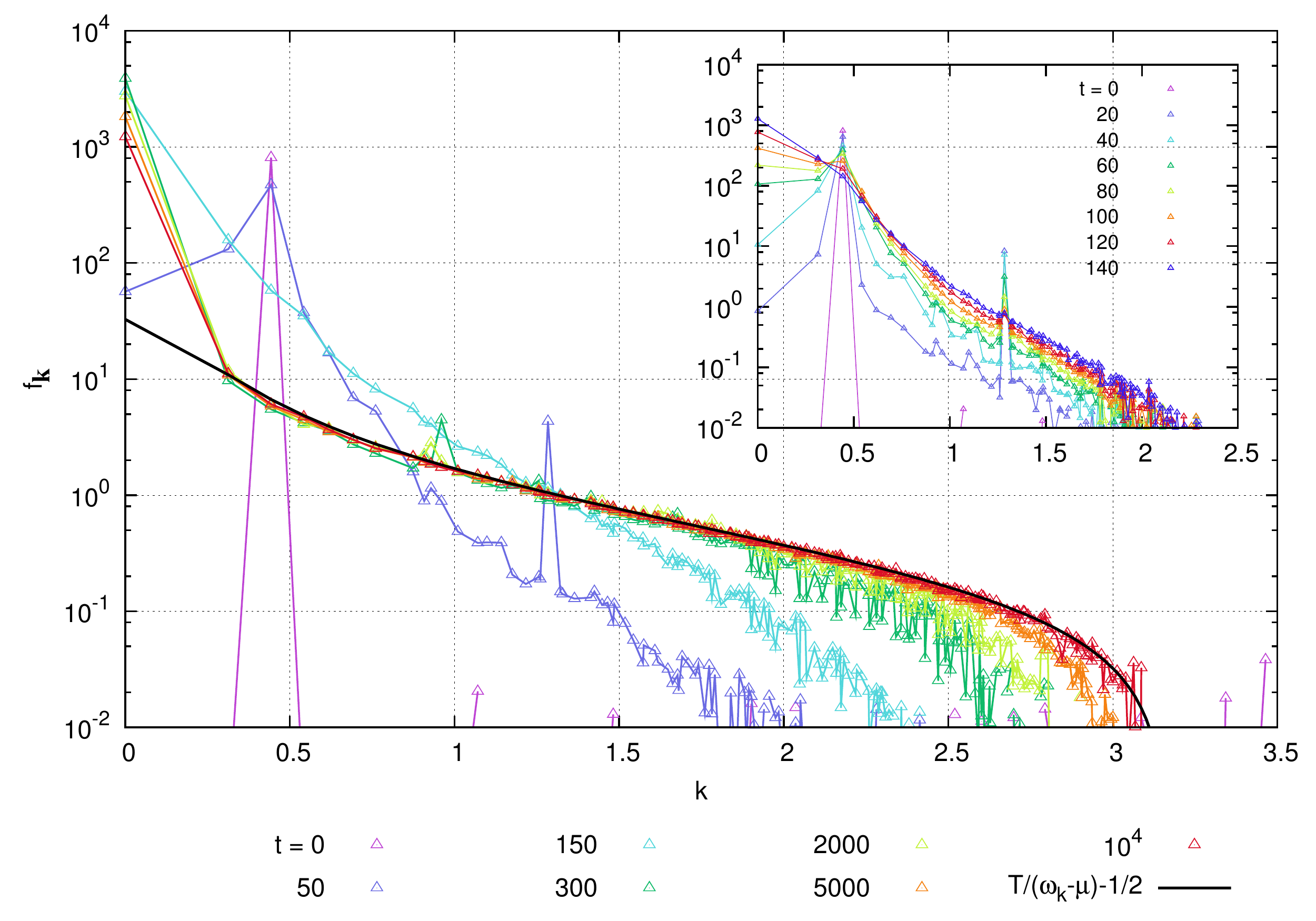}
  \hskip 5mm
  \includegraphics[width=5cm]{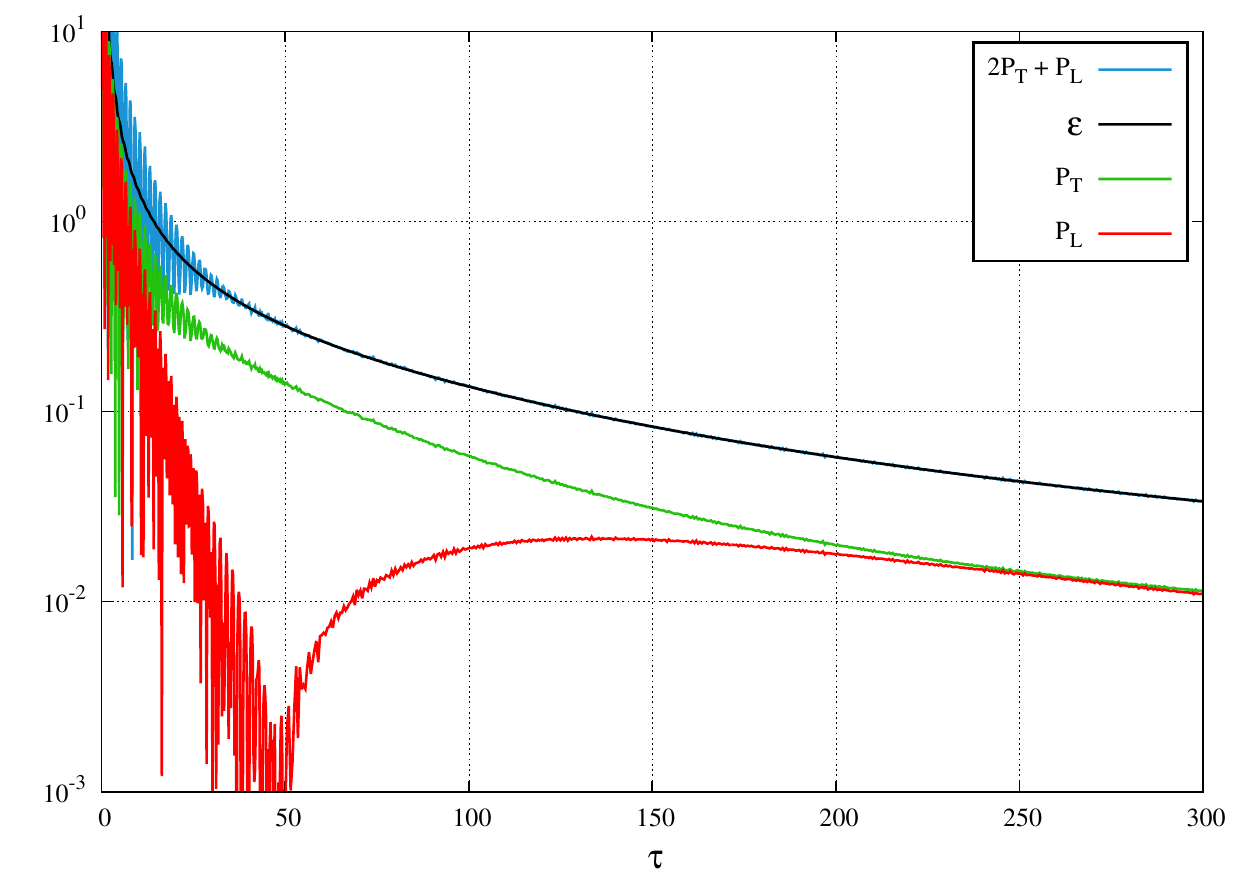}
  \caption{\label{fig:csa}CSA results in scalar field theory.}
\end{figure}
Some examples of applications \cite{Epelbaum:2011pc,Dusling:2012ig,Gelis:2013rba} of this resummation is shown in Figure
\ref{fig:csa}, in the case of a $\phi^4$ scalar field theory (known to
have unstable classical solutions due to a parametric resonance). As
one can see, the particle distribution evolves towards the equilibrium
one (but in this case, it is a classical distributions, of the form
$T/(\omega-\mu)-1/2$), and one also observes the isotropization of the
energy-momentum tensor for a longitudinally expanding system.

However, the application of the CSA with these initial conditions is
hindered by a severe sensitivity on the ultraviolet cutoff (e.g., the
lattice spacing). Indeed, the fluctuations added to the classical
field are zero-point vacuum fluctuations, and the CSA breaks the
renormalizability of the original theory by considering certain
graphs but not all of them. Within quantum field theory, the
Kadanoff-Baym equations (also known as the 2PI formalism) would
provide a renormalizable scheme that resums the relevant contributions
for thermalization. Unfortunately, its application to expanding
systems has been so far limited to a proof-of-concept study \cite{Hatta:2012gq} due to its
challenging difficulties.
\begin{figure}[htbp]
  \centering
  \includegraphics[width=6cm,height=4cm]{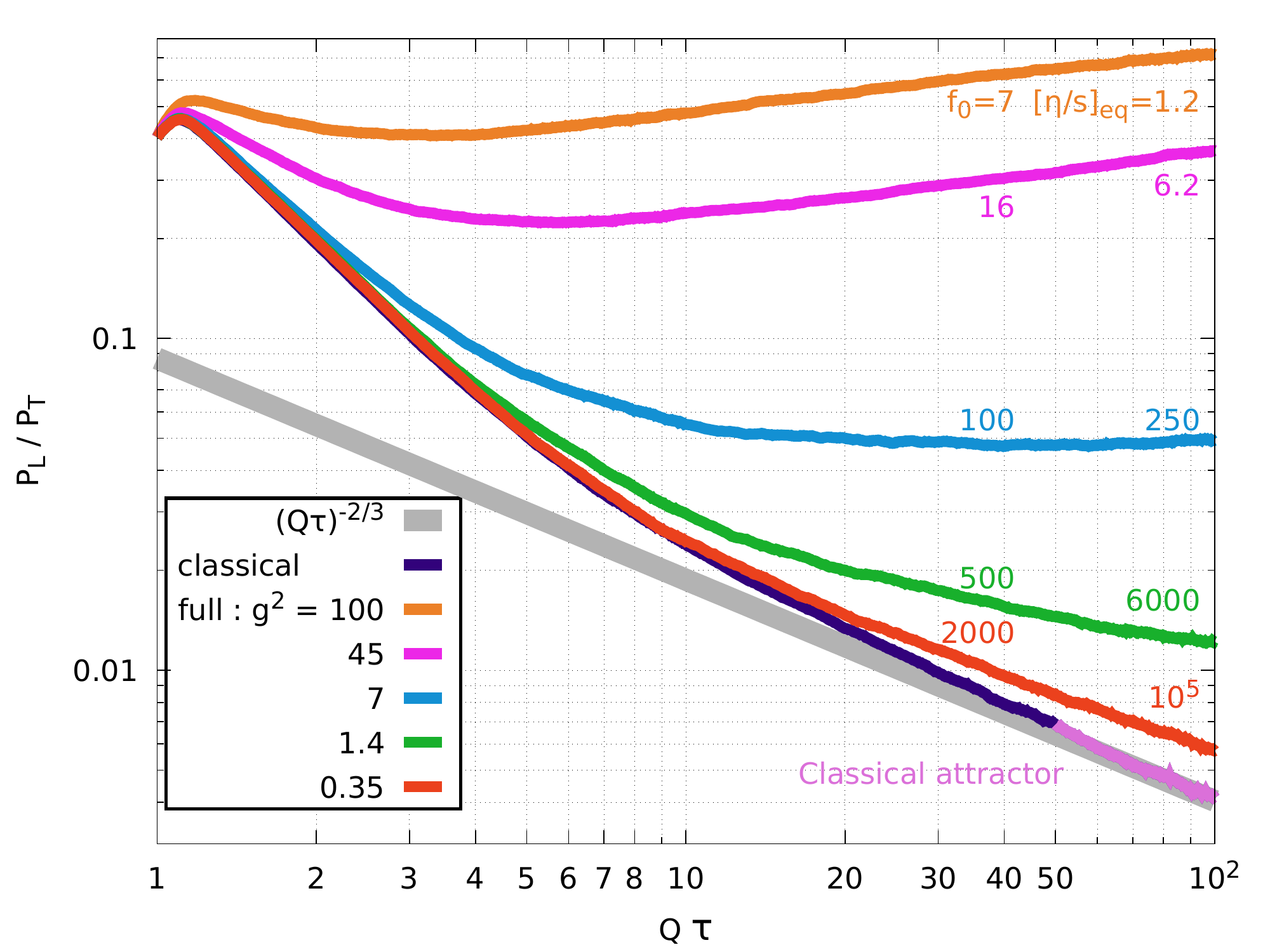}
  \hskip 5mm
  \includegraphics[width=6cm]{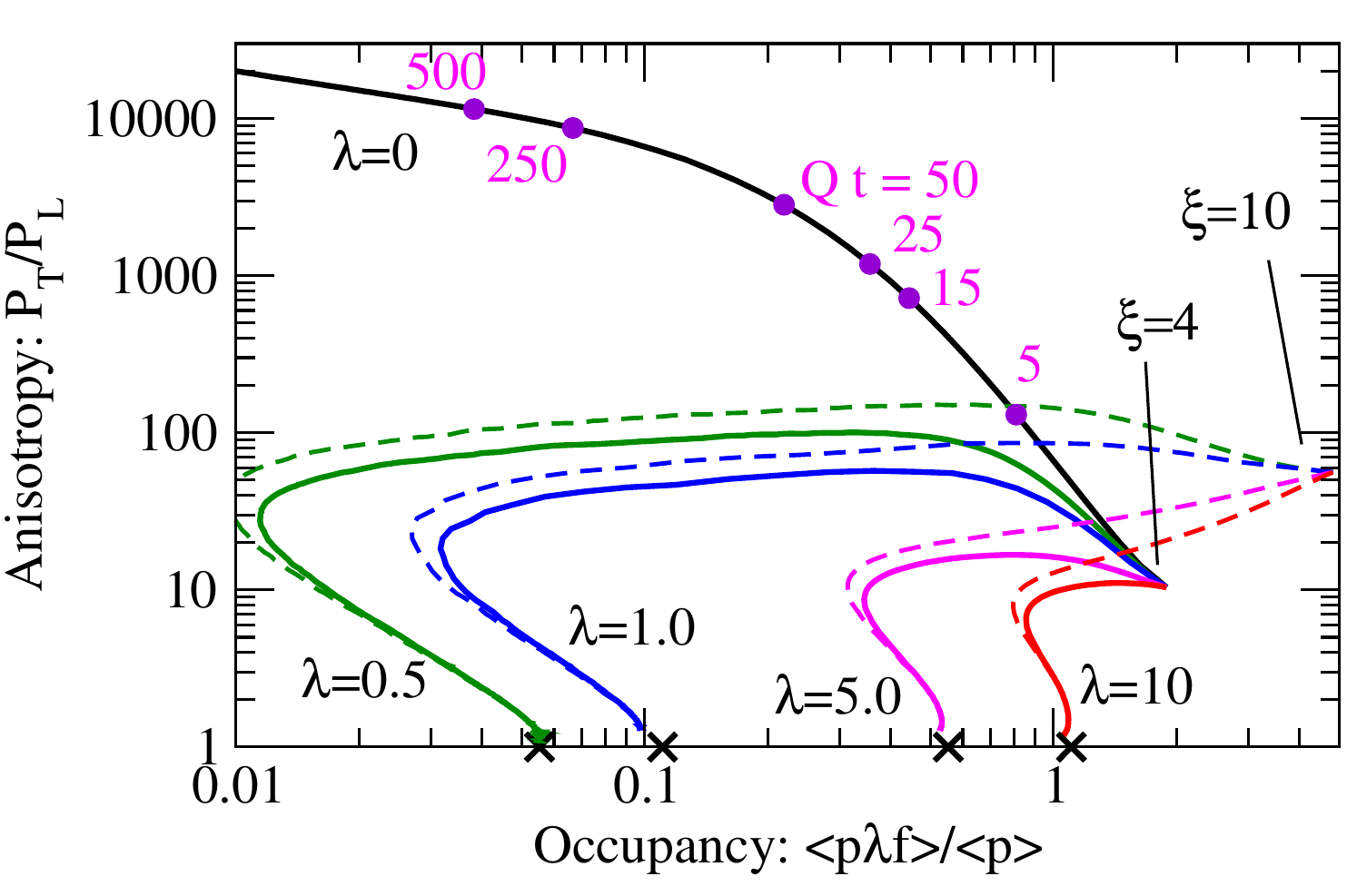}
  \includegraphics[width=6cm]{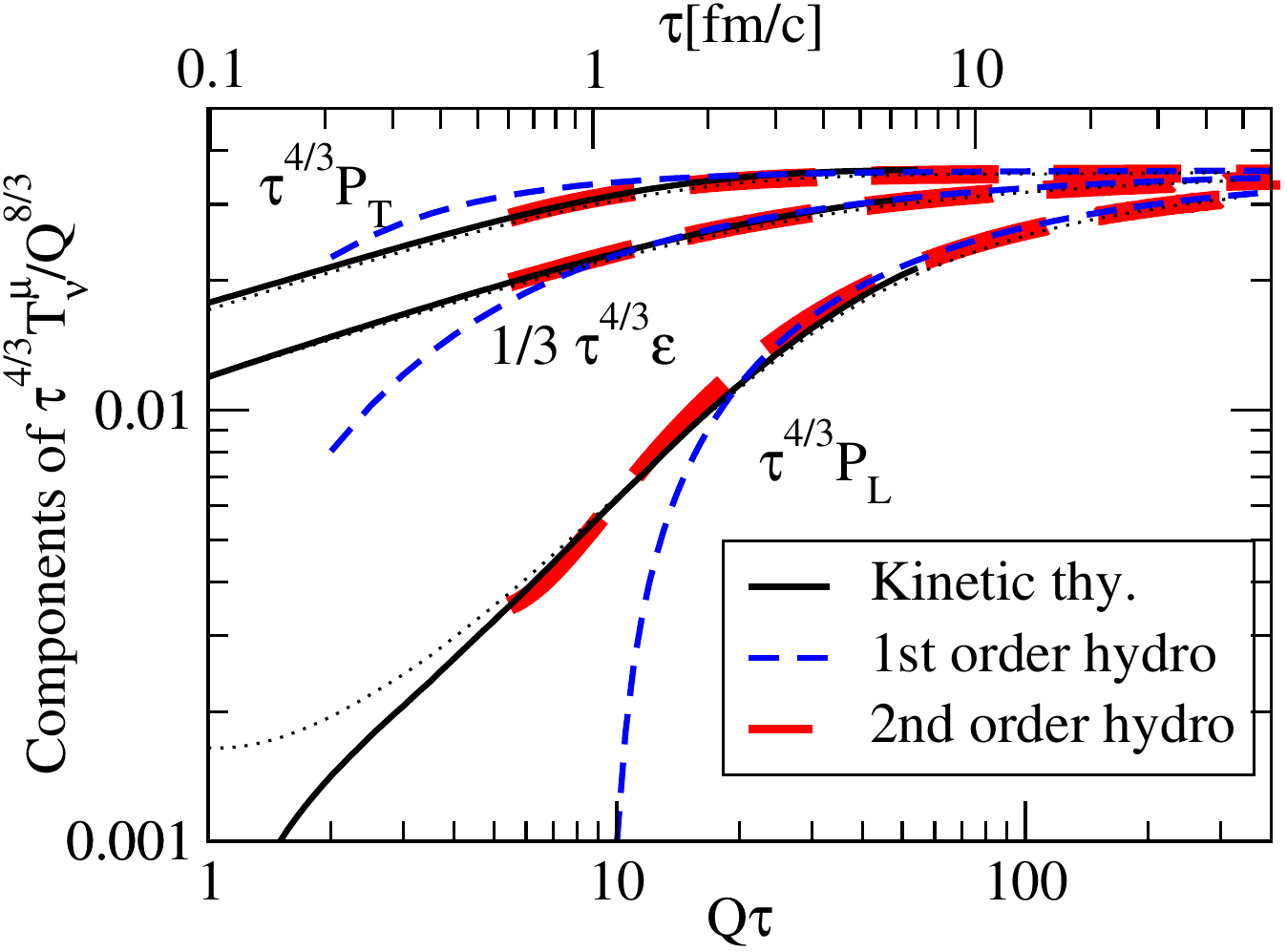}
  \hskip 5mm
  \includegraphics[width=5cm,height=4cm]{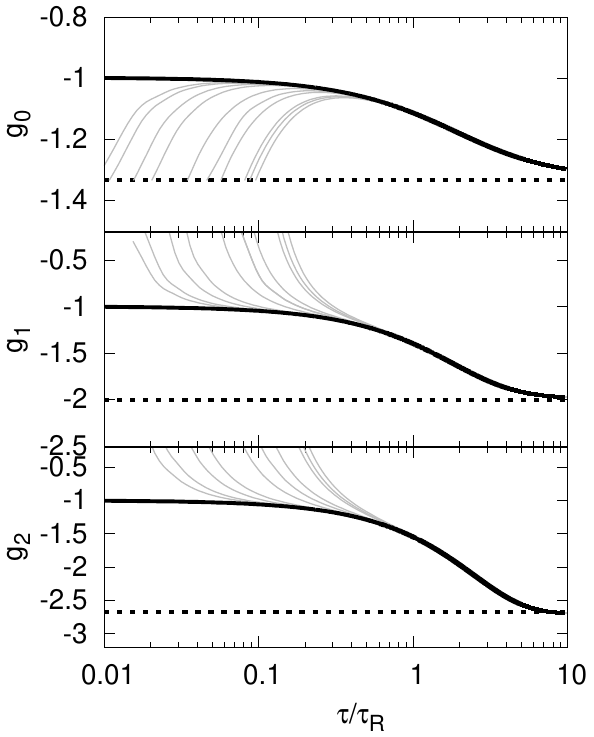}
  \caption{\label{fig:kin}Kinetic theory results. Top left:
    isotropization in an expanding system of scalar fields \cite{Epelbaum:2015vxa}. Top right:
    isotropization in Yang-Mills theory \cite{Kurkela:2015qoa}. Bottom left: comparison with
    hydrodynamics \cite{Kurkela:2015qoa}. Bottom right: the two attractors of kinetic theory
    for a longitudinally expanding system \cite{Blaizot:2017ucy}.}
\end{figure}

A much simpler to implement alternative, that requires additional
approximations (quasi-particle approximation, gradient approximation),
is kinetic theory. Various studies based on solving Boltzmann equation
have been performed, teaching the following insights:
\begin{itemize}
\item The zero-point fluctuations are crucial for the isotropization
  of the energy-momentum tensor in an expanding system (indeed, one
  may easily remove the corresponding terms in the collision kernel of
  the Boltzmann equation -- doing so prevents isotropization).
\item Isotropization is rather rapid, and the hydrodynamical behavior
  is observed significantly before full isotropization is
  achieved. Moreover, the validity of the classical approximation is
  much shorter than previously believed:
  $Q_s \tau_{\rm class} \ll \alpha_s^{-3/2}$.
\item A system undergoing longitudinal expansion has only two
  attractors: a free-streaming attractor for times small compared to
  the collision time, and an isotropic attractor when the time is
  large compared to the collision time.
\end{itemize}
Although these findings have been obtained in the approximate
framework of kinetic theory, it would be rather surprising if these
qualitative behaviors were changed significantly by going to the more
fundamental framework provided by the Kadanoff-Baym
equations. However, to ascertain these observations, it would
certainly be desirable to perform a similar study using the 2PI
framework.

\end{document}